\documentclass[longbibliography,floatfix,12pt]{iopart}

\usepackage[nospace]{cite}

  \expandafter\let\csname equation*\endcsname\relax
  \expandafter\let\csname endequation*\endcsname\relax 
    
\usepackage{bm,amsmath,amssymb,graphicx,subfigure}

\usepackage[abs]{overpic}
\newcommand{\uvc}[1]{\bm{\mathrm{\hat #1}}} 

\newcommand{\bX}{{\bf X}}

\begin{document}

\title{Rotating strings}
\author{J A Hanna}
\eads{\mailto{hanna@physics.umass.edu}}
\address{Department of Physics, University of Massachusetts, Amherst, MA 01003, U.S.A.}

\date{\today}

\begin{abstract}
Analytical expressions are provided for the configurations of an inextensible, flexible, twistable inertial string rotating rigidly about a fixed axis.  Solutions with trivial radial dependence are helices of arbitrary radius and pitch.  Non-helical solutions are governed by a cubic equation whose roots delimit permissible values of the squared radial coordinate.  Only curves coplanar with the axis of rotation make contact with it. 
\end{abstract}

The classical problems of thin strings have attracted theoretical interest since at least the seventeenth century.  They include the catenary, velaria, lintearia, and motion in a radial potential \cite{Antman05}.  Closely related to the last are the rigid rotations of a string about a fixed axis.  This problem and its generalizations, including additions of a uniform gravitational force, air drag, end masses, and tangential motion of material elements along the string, have been considered in diverse contexts by engineers, physicists, and mathematicians.  There have been abstract studies \cite{Kolodner55,Western80,Coomer01,LuningPerry84,Fusco84,Healey90,Dickey04}, as well as models of textile manufacture \cite{Mack53,Mack58,Caughey58,Gregory55,Padfield58,KothariLeaf79both,Fraser92,Zhu98,Clark98}, payload manipulation by space-, air-, and marine craft \cite{BeletskyLevin93,MankalaAgrawal05,Krupa06, Crist70,SkopChoo71,ZhuRahn98,LemonFraser01,Clark05}, jump (skipping) ropes and wind turbines \cite{Darrieus31,BlackwellReis74,MohazzabiSchmidt99,NordmarkEssen07,AristoffStone11}.  Whether thought of as tethers whirling drogues, yarn balloons, or troposkeins, these flexible bodies adopt their configurations through a simple and ubiquitous process.  Thus, it is rather surprising that a fully three-dimensional analytical treatment of the bare-bones problem remains unpublished.  This situation is likely due to the imposition of certain boundary conditions that severely restricted the solution space explored by previous researchers.  This note details solutions of the unaugmented problem, with general boundary conditions.

Given a time-dependent curve $\bX(s,t)$ parametrized by arc length $s$, the balance of inertia and line tension in an inextensible, perfectly flexible and twistable string of uniform mass density $\mu$ is described by the vector wave equation and metrical constraint
\begin{align}
	\mu\partial^2_t \bX &= \partial_s\left(\sigma\partial_s \bX\right) \, , \label{vectorwave} \\
	\partial_s \bX \cdot \partial_s \bX &= 1\, , \label{constraint}
\end{align}
where the stress $\sigma$ is a multiplier field enforcing the constraint 
\cite{EdwardsGoodyear72,Hinch94,Reeken77,Healey90,Thess99,Belmonte01,SchagerlBerger02,Preston11-2}. I will use two frames to describe the curve.  The first is the moving triad of unit vectors $(\uvc{t},\uvc{n},\uvc{b})$, defined such that
\begin{equation}
	\partial_s \left(\begin{array}{c}
			\bX\\
			\uvc{t}\\
			\uvc{n}\\
			\uvc{b}
			\end{array}\right) = \left(\begin{array}{cccc}
							0 & 1 & 0 & 0\\
							0 & 0 & \kappa & 0\\
							0 & -\kappa & 0 & \tau \\
							0 & 0 & -\tau & 0
							\end{array}\right)
	\left(\begin{array}{c}
	\bX\\
	\uvc{t}\\
	\uvc{n}\\
	\uvc{b}
	\end{array}\right) \, .
\end{equation}
The two extrinsic curvatures $\kappa$ and $\tau$ will not be important in what follows.  Velocities in this frame are represented by $\partial_t \bX \equiv T\uvc{t} + N\uvc{n}+B\uvc{b}$, with inextensibility implying $\partial_s T = \kappa N$.  The second frame is the Cartesian $(\uvc{x}, \uvc{y}, \uvc{z})$, in which I will represent the curve as
\begin{equation}
	\bX = 
	\left(\begin{array}{c}
	r(s)\cos[\omega t + \phi(s)]\\
	r(s)\sin[\omega t + \phi(s)]\\
	z(s)
	\end{array}\right) \, ,
\end{equation}
a shape consisting of material elements rotating around the $\uvc{z}$ axis with nonzero angular velocity $\omega$, with no overall flux of material along the string.  In this representation, the constraint equation \eqref{constraint} is $(\partial_s r)^2 + (\partial_s z)^2 + (r\partial_s \phi)^2 = 1$.  Finally, I will denote any constant of integration by a subscripted $c$.

Before proceeding, note that all helices, $r = R$, $\phi = Ps$, $z=\pm\left(1-R^2P^2\right)^\frac{1}{2}s$ with constant $R$ and $P$, are solutions of \eqref{vectorwave} and \eqref{constraint} bearing a uniform stress $\sigma = \frac{\mu\omega^2}{P^2}$ independent of radius.  This includes circles perpendicular to, and straight lines parallel to, the axis of rotation.  The latter have infinite stress, a pathological limit akin to the straight catenary.  

Now consider the solutions with nontrivial radial derivatives $\partial_s r$.  Projection of \eqref{vectorwave} along $\uvc{z}$ immediately gives $\sigma\partial_s z = c_1$.  So if the axial slope is zero anywhere, it is zero everywhere.
Noting that  $\partial_t T = \partial_t N = \partial_t B = 0$, projection of \eqref{vectorwave} along $\uvc{b}$ gives $T\partial_t\uvc{t}\cdot\uvc{b} = -N\partial_t\uvc{n}\cdot\uvc{b}$.  Inserting into the projections of \eqref{vectorwave} along $\uvc{t}$ and $\uvc{n}$ and multiplying these by $T$ and $N$, respectively, gives $\partial_s(\sigma T) = 0$, thus $T= \frac{c_2}{\sigma} = \frac{c_2}{c_1}\partial_s z$.  Since $T = \partial_t \bX \cdot \partial_s \bX$, $\frac{c_2}{c_1}\partial_s z = \omega r^2 \partial_s\phi$, and the azimuthal slope must also vanish everywhere or nowhere.  Observe that the curves never touch the rotational axis unless they are coplanar with it, with $T = \partial_s \phi = c_2 = 0$.  

Matching sine and cosine terms in the projection of \eqref{vectorwave} along either $\uvc{x}$ or $\uvc{y}$ leads to two more equations, one of which has already been satisfied.  The remaining equation and constraint equation may now be written in two ways.  Independent elimination of $\phi$ and $z$ gives the two pairs of equations

\begin{align}
	\partial_s\left(\frac{\partial_s r}{\partial_s z}\right) - \left(\frac{c_2}{c_1\omega}\right)^2\frac{\partial_s z}{r^3} + \frac{\mu\omega^2}{c_1}r = 0 \label{firstz} \, , \\
	\left[1+\left(\frac{c_2}{c_1\omega}\right)^2\frac{1}{r^2}\right](\partial_s z)^2 +(\partial_s r)^2 -1 = 0 \, , \label{secondz}
\end{align}
\begin{align}
	\partial_s\left(\frac{\partial_s r}{r^2\partial_s \phi}\right) - \frac{\partial_s \phi}{r} + \frac{\mu\omega^3}{c_2}r = 0 \label{firstphi} \, , \\
	\left[1+\left(\frac{c_1\omega}{c_2}\right)^2 r^2 \right](r\partial_s \phi)^2 +(\partial_s r)^2 -1 = 0 \, , \label{secondphi}
\end{align}
The limit $(c_1, \partial_s z) \rightarrow 0$ is problematic for equation pair \eqref{firstz} and \eqref{secondz}, and the limit $(c_2, \partial_s \phi) \rightarrow 0$ is problematic for equation pair \eqref{firstphi} and \eqref{secondphi}.  Subsequent manipulations will correspond to multiplications by zero and infinity in these limits, so both sets of equations must be retained and examined together.

Multiply \eqref{firstz} by $\frac{\partial_s r}{\partial_s z}$ and \eqref{firstphi} by $\frac{\partial_s r}{r\partial_s \phi}$, use \eqref{secondz} and \eqref{secondphi} to rewrite the squares of these terms, then multiply by $\partial_s z$ or $r\partial_s \phi$ and integrate to get
\begin{subequations}
\begin{align}
	\frac{1}{\partial_s z} &= c_{3\parallel} -\frac{\mu\omega^2}{2c_1}r^2 \, , \\
	\frac{1}{\omega r^2\partial_s \phi} &= c_{3\perp} -\frac{\mu\omega^2}{2c_2}r^2 \, .
\end{align}
\end{subequations}
Squaring these and using \eqref{secondz} and \eqref{secondphi} again to eliminate $\partial_s z$ and $r\partial_s \phi$ gives
\begin{subequations} \label{dsr}
\begin{align}
	(\partial_s r)^2 &= 1 - \frac{4}{\left(2c_1c_{3\parallel}-\mu\omega^2r^2\right)^2}\left(c_1^2 + \frac{c_2^2}{\omega^2 r^2}\right) \label{dsrz} \, , \\
	(\partial_s r)^2 &= 1 - \frac{4}{\left(2c_2c_{3\perp}-\mu\omega^2r^2\right)^2}\left(c_1^2 + \frac{c_2^2}{\omega^2 r^2}\right) \, . \label{dsrphi} 
\end{align}
\end{subequations}
The limit $(c_1, c_2) \rightarrow 0$ is the straight radial line $r=\pm s$ with $(z, \phi)$ constant.  Taking $c_2 \rightarrow 0$ in \eqref{dsrz} leads to Mack's equation \cite{Mack58} for strings coplanar with the axis of rotation, while taking $c_1 \rightarrow 0$ in \eqref{dsrphi} leads to Fusco's equation \cite{Fusco84} for strings in the plane perpendicular to the axis of rotation.  At first glance, these limits only give sensical results if they are taken for the correct equation.  However, the products of integration constants in the denominators actually represent the same physical quantity, and equations \eqref{dsr} are, in fact, the same equation.  Let's thus define $c_3 \equiv c_1c_{3\parallel} =  c_2c_{3\perp}$, and take stock of our constants of integration.  They may be expressed in terms of three vectors, namely the velocity, stress vector, and unit axis of rotation:
\begin{align}
	\sigma \partial_s z =& \; c_1   = \sigma\uvc{t}\cdot\uvc{z} \, , \label{c1} \\
	\sigma \omega r^2 \partial_s \phi =& \; c_2  = \sigma\uvc{t}\cdot\partial_t\bX \, , \label{c2} \\
	\sigma + \frac{\mu\omega^2r^2}{2} =& \; c_3  = \left(\sigma\uvc{t}\cdot\sigma\uvc{t}\right)^\frac{1}{2} + \frac{\mu}{2}\partial_t\bX\cdot\partial_t\bX \, . \label{c3}
\end{align}
These expressions allow recovery of the helical solutions after setting $\partial_s r = 0$ in \eqref{dsr}.  The first two constants represent measures of the axial slope and azimuthal slope, while the third is suggestive of an energy.  One could divide these by the mass density and suitable powers of frequency, and then normalize by a relevant length scale such as the total length of string involved in the situation of interest. Presently, there is no such relevant scale, and it seems most sensible to use $c_3$ to provide the length in the normalizing factor.  Define the nondimensional arc length $\hat{s} \equiv s \left( \frac{\mu\omega^2}{2c_3} \right)^\frac{1}{2}$ and two parameters $\alpha \equiv c_1 \frac{1}{c_3}$ and $\beta \equiv c_2  \left(\frac{\mu}{2c_3^3} \right)^\frac{1}{2}$, then substitute $u \equiv r^2 \frac{\mu\omega^2}{2c_3}$ in \eqref{dsr} and fiddle a bit to obtain
\begin{equation}\label{radiusflow}
	\pm \partial_{\hat{s}} u = \frac{2}{1 - u}\left[ u^3 - 2 u^2 + (1-\alpha^2)u-\beta^2 \right]^{\frac{1}{2}} \, .
\end{equation}
From this, one may write an implicit solution in terms of an elliptic integral \cite{ByrdFriedman54},
\begin{equation}\label{implicit}
	\pm \hat{s} = \frac{1}{2} \int^u \!\!\! d\tilde{u} \, \frac{1 - \tilde{u}}{\left[\tilde{u}^3-2\tilde{u}^2+(1-\alpha^2)\tilde{u}-\beta^2\right]^{\frac{1}{2}}} \, .
\end{equation}
Expressions for $\hat{z} \equiv z \left( \frac{\mu\omega^2}{2c_3} \right)^\frac{1}{2}$ and $\phi$ are
\begin{align}
	\pm \hat{z} &= \int^{\hat{s}} \!\!\! d\tilde{s} \, \frac{\alpha}{1 - u(\tilde{s})} \, , \label{zimplicit}\\
	\pm \phi &= \int^{\hat{s}} \!\!\! d\tilde{s} \, \frac{\beta}{u(\tilde{s})\left[1 - u(\tilde{s})\right]} \, .  \label{phiimplicit}
\end{align}

If the physical assumptions $\sigma > 0$ and $\mu > 0$ are made, then relations \eqref{c1} and \eqref{c3} imply $c_3 > 0$ and $c_3^2 > c_1^2$.  Thus, the cubic coefficients in \eqref{radiusflow} or \eqref{implicit} are such that $-2 <0$, $1-\alpha^2>0$, and $-\beta^2<0$.  Use of ``Descartes's rule of signs'' shows that there are no real negative roots of the cubic--- there can be three or one real positive roots, depending on the relative magnitudes of the parameters.  If $\beta=0$, solutions are coplanar with the axis of rotation, one root is zero and the other two are $1 \pm \alpha$.  If $\alpha=0$, solutions are perpendicular to the axis of rotation, and there will be three real roots when $\beta^2 \le \frac{4}{27}$, one real root otherwise.  In the general case, it is a tedious but straightforward process to find all three roots in terms of the slope parameters $\alpha$ and $\beta$, and explicitly reconstruct the parameters in terms of the roots.

For non-helical solutions to exist, both the cubic and $u$ must be non-negative.  With three real roots, there are two branches of solutions with non-constant radius: an inner branch where $u$ is larger than the smallest root but smaller than the other two, and an outer branch where $u$ is larger than all three roots.  With one real root, there is only an outer branch, where $u$ is larger than the real root.  Note, however, that the outer branch of the cubic will correspond to a configuration that is concave outward from the axis of rotation, and thus in a state of compressive stress.  This may be easily seen algebraically for the coplanar case, where the outer branch corresponds to $u > 1 + | \alpha | > 1$, which implies through \eqref{c3} that the stress is compressive.  A compressive stress in \eqref{vectorwave} corresponds to inherently unstable dynamics that are essentially impossible to observe in real life.  So the important physical solutions are those of the inner branch, which only exists for sufficiently small relative values of $\beta$ such that the cubic discriminant is non-negative.  This bound on one of the parameters was previously noted, through a different argument, in Fusco's study of periodic, perpendicular-planar strings \cite{Fusco84}.  The inner branch coplanar solutions are the only spatial curves that touch the axis of rotation.  Explicit parametrizations of these curves are known, with $z$ expressed as a function of $r$ \cite{MohazzabiSchmidt99,NordmarkEssen07}.

\begin{figure}[here]
\subfigure{
	\begin{overpic}[width=3in]{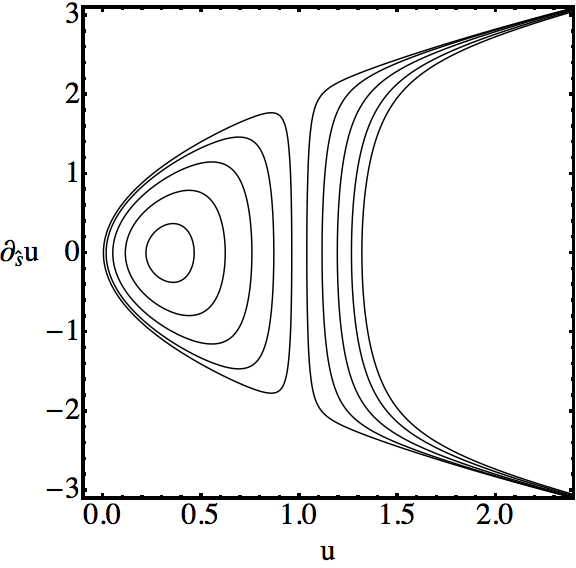}\label{first}
	\put(50,180){\Large{\subref{first}}}
	\end{overpic}
	}
\subfigure{
	\begin{overpic}[width=3in]{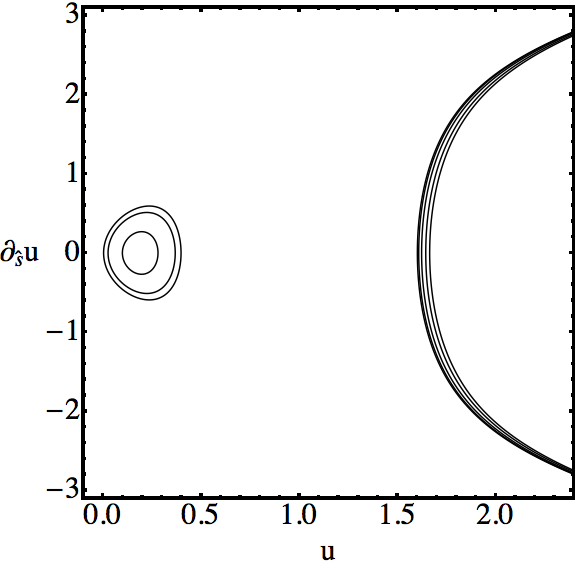}\label{second}
	\put(50,180){\Large{\subref{second}}}
	\end{overpic}
	}\\
\subfigure{
	\begin{overpic}[width=3in]{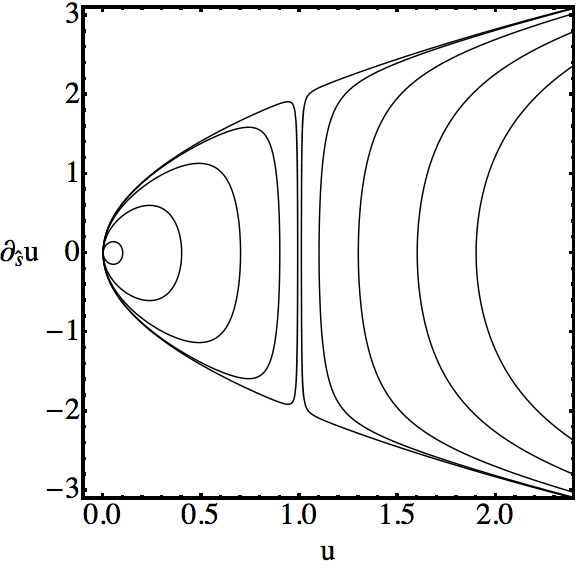}\label{third}
	\put(50,180){\Large{\subref{third}}}
	\end{overpic}
	}
\subfigure{
	\begin{overpic}[width=3in]{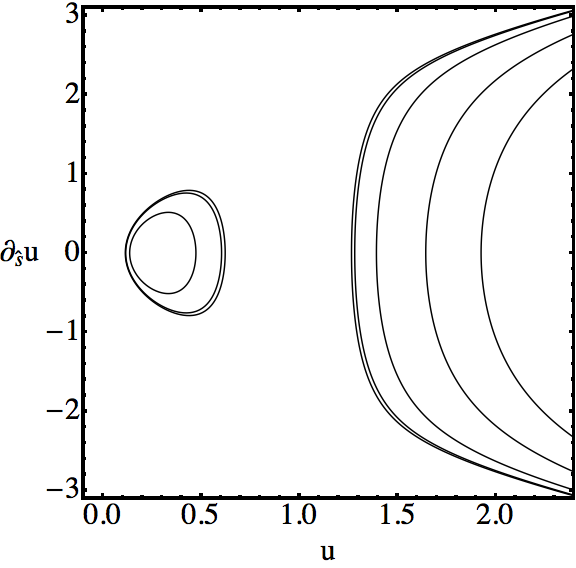}\label{fourth}
	\put(50,180){\Large{\subref{fourth}}}
	\end{overpic}
	}
\caption{ Phase portraits for $u$, obtained from equation \eqref{radiusflow}.  All curves lie inside the parabola $(\partial_{\hat{s}}u)^2=4u$ and intersect the $u$ axis at the roots of the cubic equation $u^3 - 2u^2 + (1-\alpha^2)u - \beta^2=0$\,.  Increasing either $\alpha$ or $\beta$ shrinks the trajectories.  \subref{first} The perpendicular-planar case $\alpha=0$, $\frac{27}{4}\beta^2 = (0.01, 0.1, 0.3, 0.6, 0.9)$.  The curves do not touch the $\partial_{\hat{s}}u$ axis.  \subref{second} $\alpha=0.6$, $\frac{27}{4}\beta^2 = (0.01, 0.1, 0.3, 0.6, 0.9)$\,,  \subref{third} The coplanar case $\alpha=(0.01, 0.1, 0.3, 0.6, 0.9)$\,, $\frac{27}{4}\beta^2 = 0$.  The curves touch the $\partial_{\hat{s}}u$ axis.  \subref{fourth} $\alpha=(0.01, 0.1, 0.3, 0.6, 0.9)$, $\frac{27}{4}\beta^2 = 0.6$\,. }
\label{phaseportraits}
\end{figure}

Much of this behavior is more easily seen in phase portraits for $u$, obtained from equation \eqref{radiusflow}.  Four such portraits, displaying different choices of slope parameters, are shown in Figure \ref{phaseportraits}.  They consist of a closed inner tensile lobe and open outer compressive lobe.  Increasing either parameter results in smaller lobes.  The lobes merge into the line $u=1$ and a single parabola $(\partial_{\hat{s}}u)^2=4u$ when both $\alpha$ and $\beta$ go to zero; this is the straight radial line solution.  The definition of the $s$ coordinate as arc length constrains all trajectory curves to lie inside this parabola.  The curves intersect the $u$ axis at the roots of the cubic.  Figure \ref{first}, which looks like a fish, shows perpendicular-planar solutions.  The inner lobe curves do not touch the $\partial_{\hat{s}}u$ axis for nonzero $\beta$.  For ($\alpha=0$, $\beta^2=0$) the roots are (0, 1, 1), and for ($\alpha=0,\beta^2=\frac{4}{27}$) the roots are ($\frac{1}{3}$, $\frac{1}{3}$, $\frac{4}{3}$).  Increasing $\alpha$ as in Figure \ref{second} shrinks the lobes, the inner lobe shifting towards the $\partial_{\hat{s}}u$ axis; the two smallest inner lobe curves have disappeared.  Figure \ref{third} shows coplanar solutions.  The inner lobe curves touch the $\partial_{\hat{s}}u$ axis, and do so with a generically nonzero slope $\partial_s r$, although $\partial_{\hat{s}}u = 0$ when $u=0$.  For ($\alpha=1$, $\beta^2=0$) the roots are (0, 0, 2).  Increasing $\beta$ as in Figure \ref{fourth} shrinks the lobes, the inner lobe shifting off of the $\partial_{\hat{s}}u$ axis; the two smallest inner lobe curves have disappeared.
    
\begin{figure}[here]
\subfigure{
	\begin{overpic}[height=3.0in]{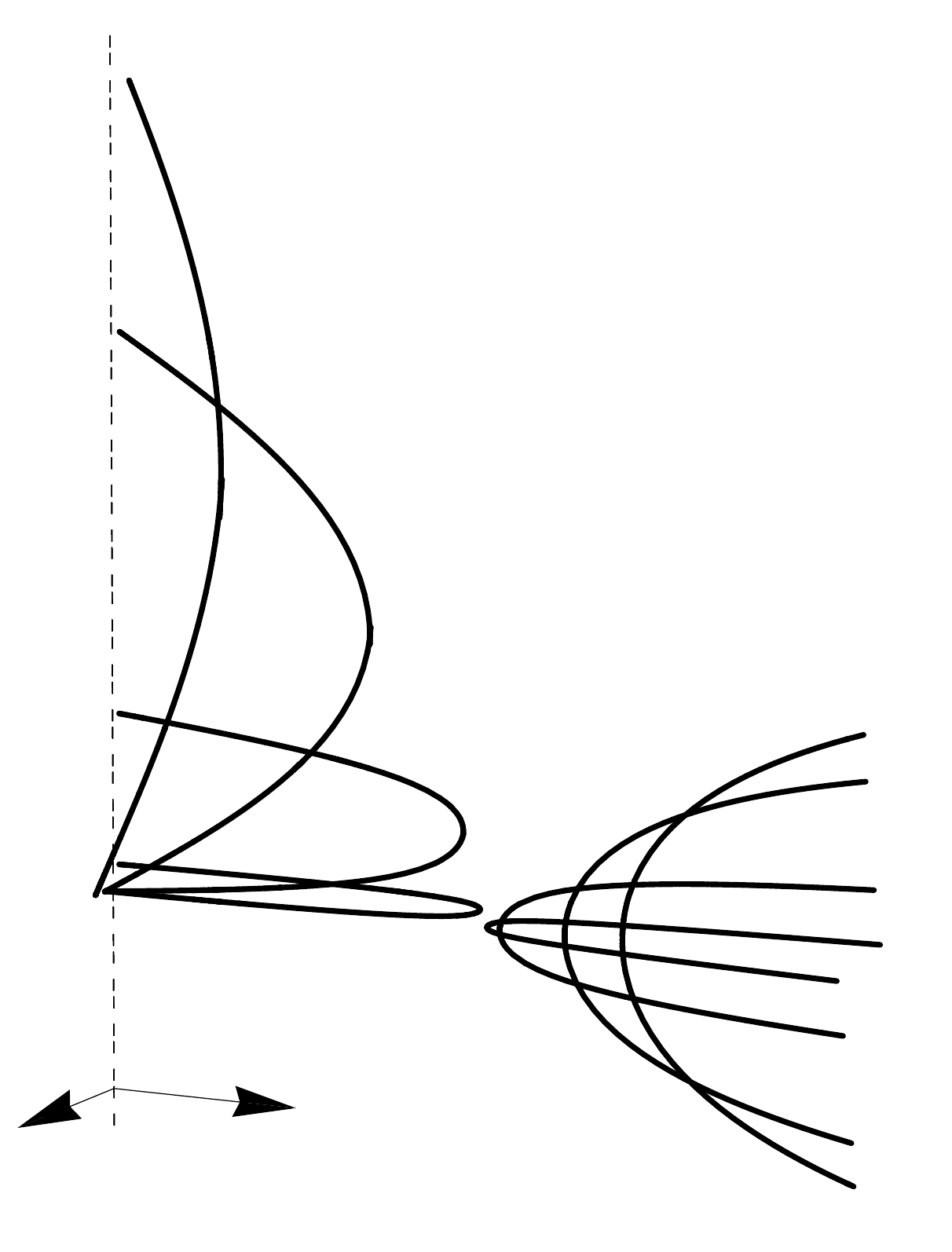}\label{smallbeta1}
	\put(100,180){\Large{\subref{smallbeta1}}}
	\end{overpic}
	}
\subfigure{
	\begin{overpic}[height=3.0in]{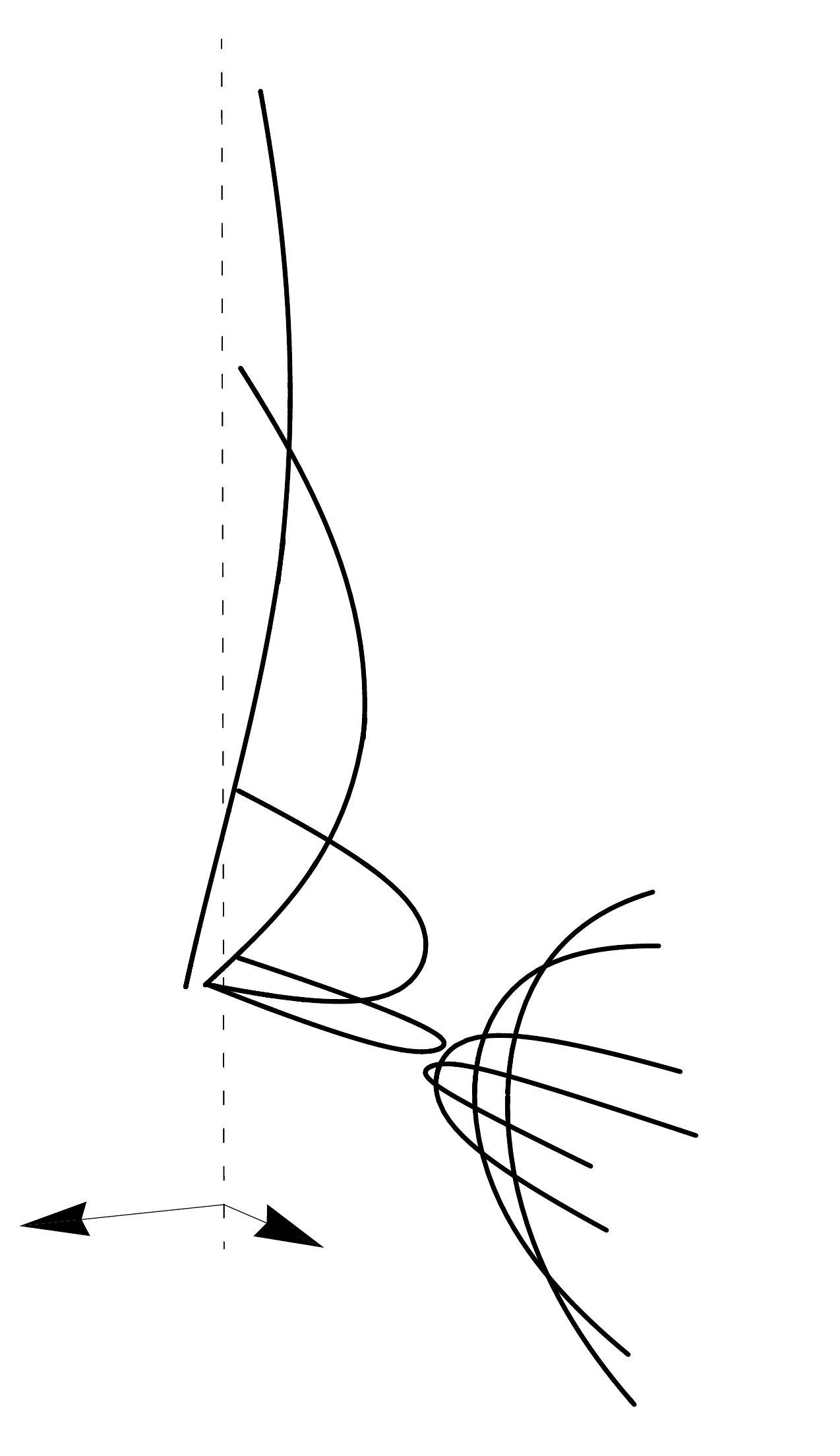}
	\end{overpic}
	}
\subfigure{
	\begin{overpic}[height=3.0in]{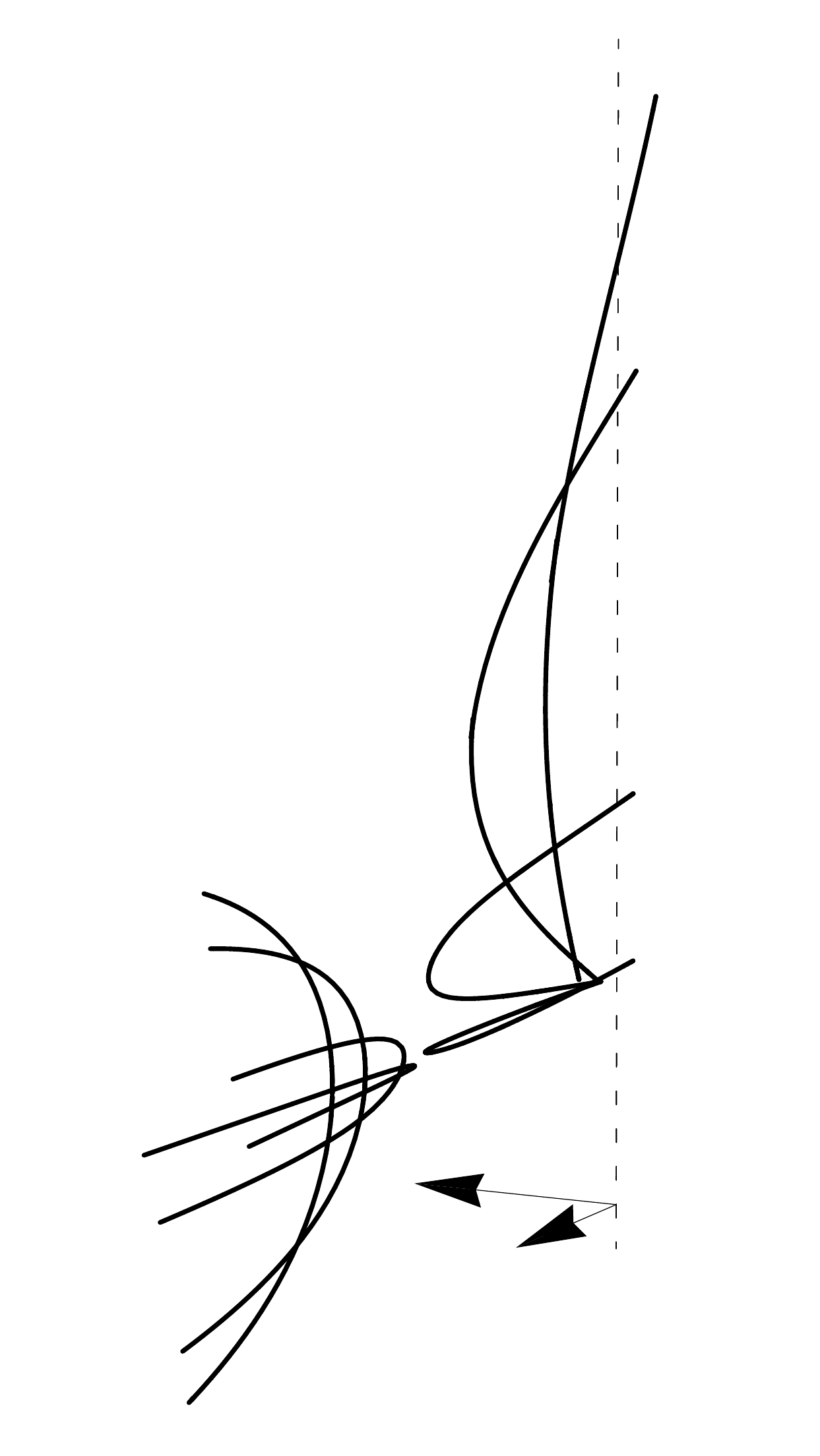}
	\end{overpic}
	}\\
	\addtocounter{subfigure}{-2}
\subfigure{
	\begin{overpic}[height=3.0in]{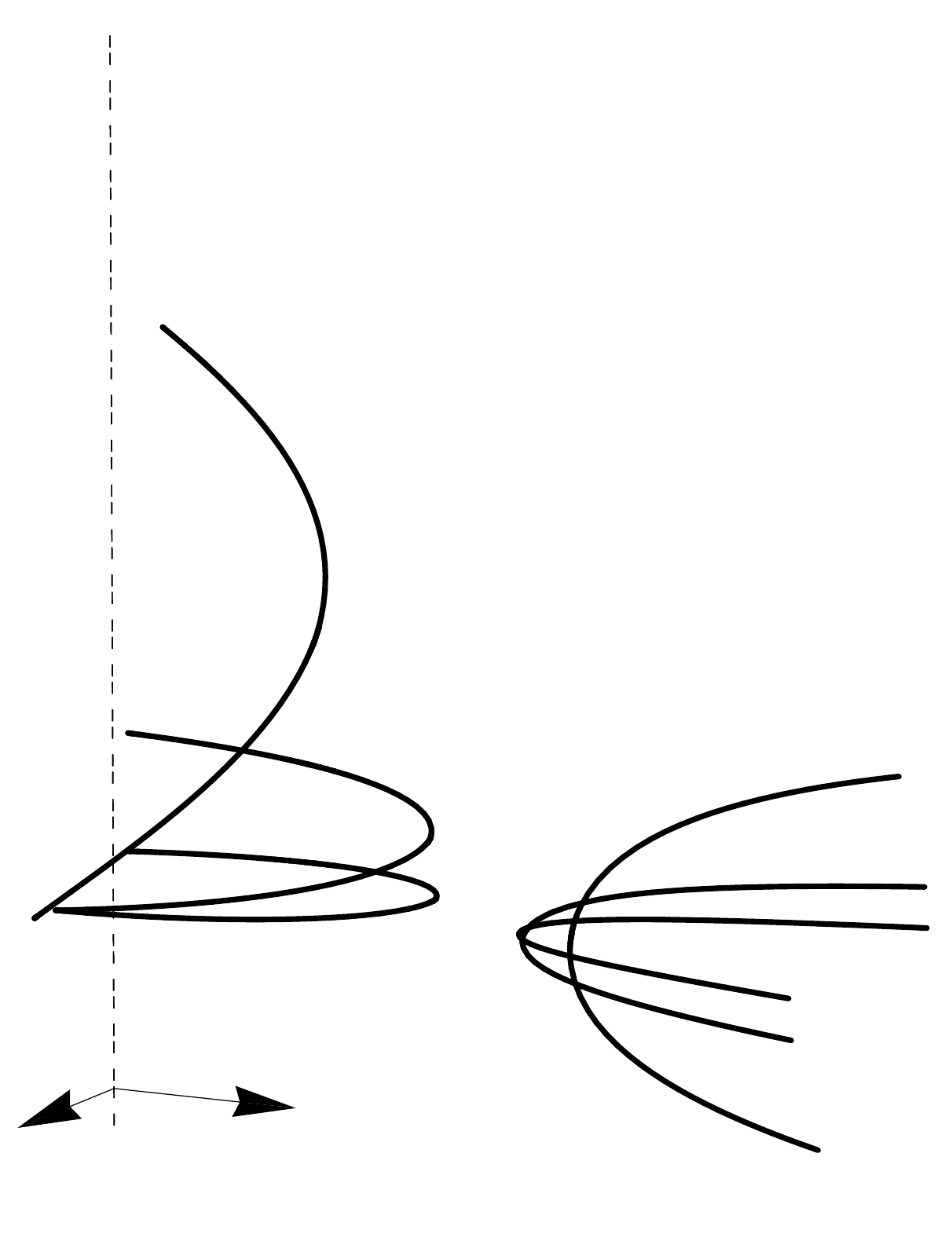}\label{mediumbeta1}
	\put(100,180){\Large{\subref{mediumbeta1}}}
	\end{overpic}
	}
\subfigure{
	\begin{overpic}[height=3.0in]{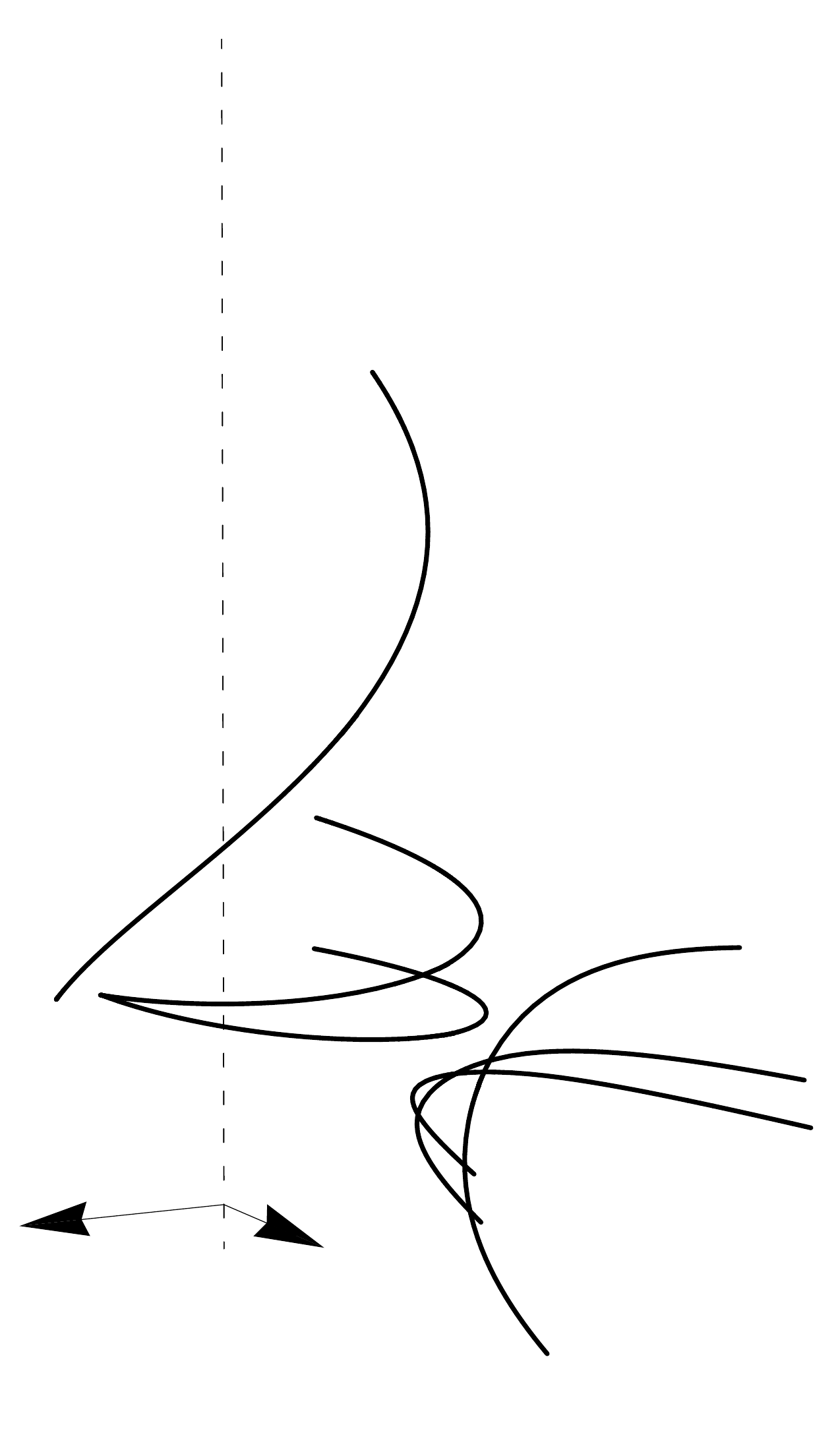}
	\end{overpic}
	}
\subfigure{
	\begin{overpic}[height=3.0in]{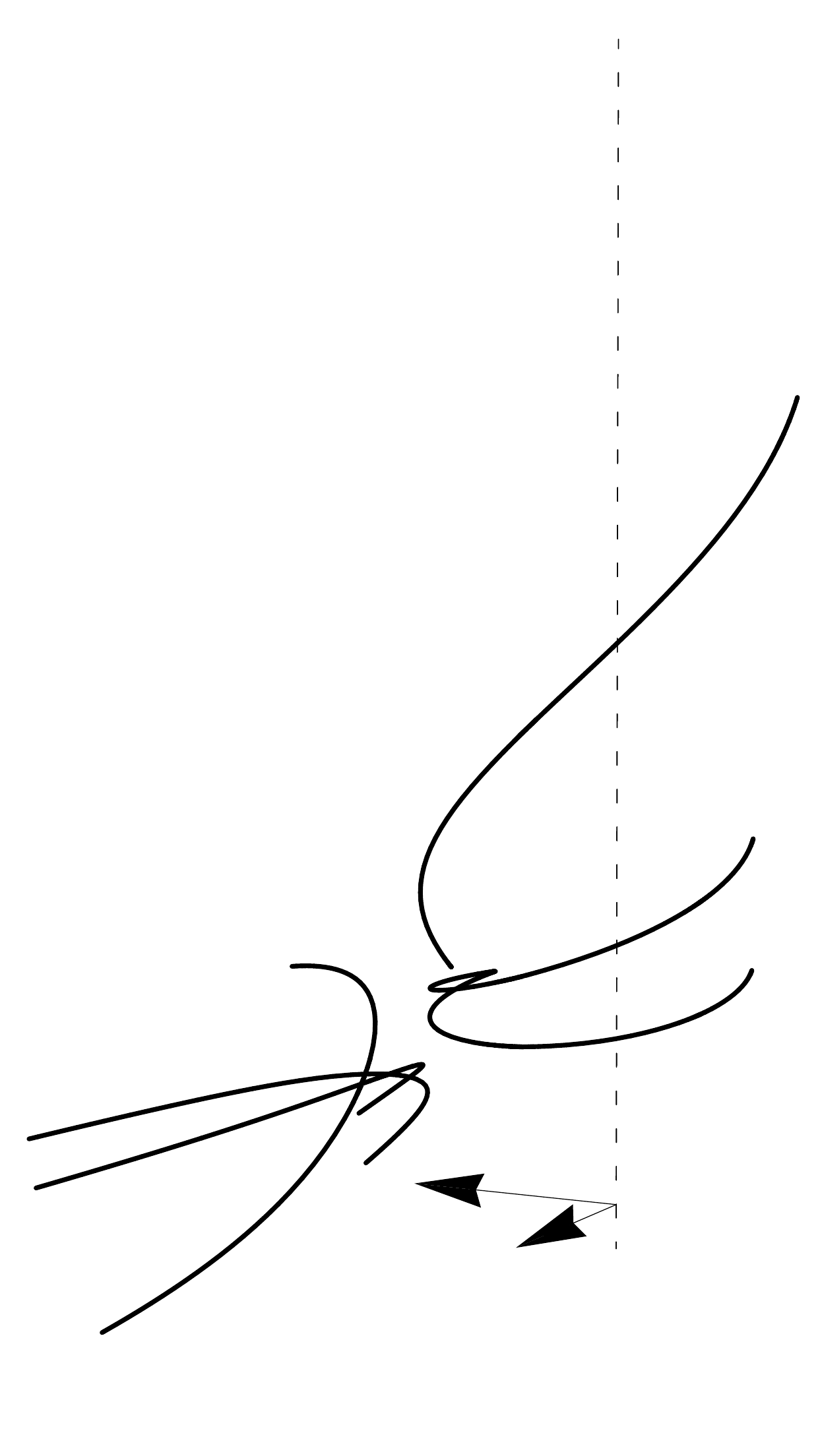}
	\end{overpic}
	}\\
	\addtocounter{subfigure}{-2}
\caption{ A menagerie of non-planar jumpropica.  Vertical dotted lines represent the rotational axis, and orthogonal arrowed lines are shown to guide the eye.  \subref{smallbeta1} Three views of \mbox{$\alpha = (0.01, 0.1, 0.5, 0.9)$, $\frac{27}{4}\beta^2 = 0.01$ .}  \subref{mediumbeta1} Three views of \mbox{$\alpha = (0.01, 0.1, 0.5)$, $\frac{27}{4}\beta^2 = 0.5$ .} }
\label{menagerie}
\end{figure}

A few spatial curves are shown in Figure \ref{menagerie}.  They resemble jump ropes for small $\beta$ and flower petals for small $\alpha$, and may be extended indefinitely to any length.  Any curve's chirality may be reversed to produce another solution, and rotation may occur in either sense, as changing the sign of $\omega$ does not affect the equations for $r$, $\phi$, and $z$.

Consider applying boundary conditions to an end of the string, at a given radius corresponding to a nondimensional squared radius $u_0$ that lies within one of the ranges allowed by the cubic.  Here, $\frac{\alpha}{1-u_0}$ and $\frac{\beta}{u_0\left(1-u_0\right)}$ are the values of the axial and azimuthal slopes, respectively, and one must apply a nondimensional stress $\frac{\sigma}{c_3} = 1 - u_0$ along the tangent.  Requiring periodicity of $\bX$ restricts the problem to the perpendicular plane \cite{Fusco84}, while requiring a vanishing radius yields only the coplanar solutions.  It is likely that a vanishing stress condition, such as would be found at a free end, is similarly restrictive for the corresponding problem in the presence of gravity.  Imposition of one or more of these conditions at the outset of considering this problem, or an augmented form of it, may thus lead to neglect of three-dimensional solutions.

Although the slope parameters provide a natural description of the shapes, they do not correspond to experimentally imposable boundary conditions.  Perfectly flexible strings cannot be clamped; one can only fix some selection of radial, axial, and azimuthal positions of the two ends, by attaching them to rotating rigid supports that may have one or more degrees of freedom in their positions and orientations.  These boundary conditions are connected to the slopes through integral relationships, and there is an additional global constraint on the total length of the string, all of which is implicitly encoded in equations (\ref{radiusflow}-\ref{phiimplicit}).  With length and some positions fixed by the scientist, the slopes and stress will be chosen by the physical system.  Let us consider only the tensile solutions, the closed orbits in $u$-$\partial_{\hat{s}}u$ space.  Choosing two radial positions within an appropriate region gives us, generically, a two-parameter continuous family of orbits on which solutions may be found.  By subsequently choosing two additional constraints, we select a discrete family of solutions through the global relationships (\ref{radiusflow}-\ref{phiimplicit}).  In practice, one of the additional constraints is likely to be total length.  The other could be total axial distance between the supports, or total azimuthal angular difference modulo $2\pi$.  Although the axial and azimuthal positions increase monotonically during traversal of an orbit, multiple solutions are still a possibility because we cannot fix the number of turning points of the radius \footnote{If the supports are fixed, then transition between states of different azimuthal winding number by a continuous deformation can only occur if the string can penetrate the axis of rotation.  This fact may prove to be an experimental nuisance.  However, the winding number of the orbit, which is the number of pairs of turning points of the radius, can differ from the winding number of the azimuthal angle.}.  

This short study may be expanded in several ways.  A logical next step, especially if one is concerned with industrial yarn problems, is to add a constant tangential velocity $T_0$ to the string so that $\partial_t\bX \rightarrow \partial_t\bX + T_0\uvc{t}$ and material elements move along the rigid shape.  From Mack's analysis \cite{Mack58}, it appears that this will add terms proportional to $r^2$ inside the rightmost parentheses in equations \eqref{dsr}.  This will not change the order of equation \eqref{radiusflow}, but should add new solutions \emph{via} a third parameter in the equation that represents the relative strength of Coriolis forces.  This string problem contrasts with those that occur in inertial, non-rotating frames, where adding tangential motion leads only to centripetal forces.  Unlike Coriolis forces, centripetal forces are balanced merely by a change in the stress \cite{Routh55}, which does not modify the shapes of the solutions, but might stabilize some portion of those that would otherwise bear compressive stresses.  

The addition of a uniform gravitational force along the axis of rotation will likely have the nontrivial effect of destroying the periodicity of $r$, as the solutions of the corresponding geometrically linearized problem are Bessel functions.  An interesting set of non-rigid solutions to look for, perhaps using the rigid helical solutions as a starting point, are the ``subharmonic modes of rotation'' observed by Caughey, ``in which the fixed end of the chain performs an integral number of rotations in the time that the free end takes to make one complete revolution'' \cite{Caughey58}.  

Recent numerical work by Aristoff and Stone \cite{AristoffStone11} suggests that the addition of air drag to a coplanar solution, while keeping the boundary condition $r=0$, leads to curvature singularities at the radial maxima.  The effect of drag on rotating thin structures has direct relevance to the design of wind and water turbines.  It is curious that, while the patent for the Darrieus turbine \cite{Darrieus31} cites a rotating rope's shape as that which will minimize the bending moments on a slender blade, the helical shape of this turbine's progeny, the Gorlov turbine \cite{Gorlov95}, was chosen for other reasons, apparently unrecognized as another choice that would satisfy the same criterion.

Finally, those forbidden regions of phase space which correspond to negative $u$ might be explored by interpreting $\omega$ as imaginary, and considering a scattering problem for strings in a cylindrical potential.

\section*{Acknowledgments}

I thank J. Aristoff for sharing a preprint and pointing me towards some references, and C. D. Santangelo for helpful input and support through U.S. National Science Foundation grant DMR 0846582.  These results were available free of peer review on the arXiv in 4/2012.


\section*{References}

\bibliographystyle{unsrt}

\end{document}